\documentclass{article}
\usepackage{amsfonts}
\usepackage{amsmath}

\begin{document}

\title{Splitting the Dirac equation: the case of longitudinal potentials}
\author{Andrzej Okninski \\
Physics Division, Department of Management and Computer\\
Modelling, Politechnika Swietokrzyska, Al. 1000-lecia PP 7, \\
25-314 Kielce, Poland}
\maketitle

\begin{abstract}
Recently, we have demonstrated that some subsolutions of the free
Duffin-Kemmer-Petiau and the Dirac equations obey the same Dirac equation
with some built-in projection operators.

In the present paper we study the Dirac equation in the interacting case. \
It is demonstrated that the Dirac equation in longitudinal external fields
can be also splitted into two covariant subequations.
\end{abstract}

\section{\label{Intro}Introduction}

Recently, several supersymmetric systems, concerned mainly with anyons in
2+1 dimensions \cite
{Jackiw1991,Plyushchay1991,Horvathy2010a,Horvathy2010b,Horvathy2010c} as
well as with the 3+1 dimensional Majorana-Dirac-Staunton theory \cite
{Horvathy2008}, uniting fermionic and bosonic fields, have been described. 
Furthermore, bosonic symmetries of the Dirac equation have been found  in 
the massless \cite{Simulik1998} as well as in the massive case \cite{Simulik2011}. 
Our results derived lately fit into this broader picture. We have
demonstrated that some subsolutions of the Duffin-Kemmer-Petiau (DKP) and
the Dirac equations obey the same Dirac equation with some built-in
projection operators \cite{Okninski2011}. We shall refer to this equation as
supersymmetric since it has bosonic (spin $0$ and $1$) as well as fermionic
(spin $\frac{1}{2}$) degrees of freedom. In the present paper we extend our
results to the Dirac equation in external longitudinal fields.

The paper is organized as follows. In Section \ref{Dirac} the Dirac equation
as well as conventions and definitions used in the paper are described
shortly. In the next Section the Dirac equation in the interacting case is
splitted into two subsolutions equations -- Dirac equations with built-in
projection operators. We demonstrate that this is possible in the case of
longitudinal potentials. In Section \ref{Separation} variables are separated
in the subsolutions equations to yield $2D$ Dirac equations in $
\left( x^{0},x^{3}\right) $ subspace and $2D$ Pauli equations
in $\left( x^{1},x^{2}\right) $ subspace and in the last Section we discuss
the emerging picture.

\section{\label{Dirac}The Dirac equation}

In what follows tensor indices are denoted with Greek letters, $\mu =0,1,2,3$%
. We shall use the following convention for the Minkowski space-time metric
tensor: $g^{\mu \nu }=$ \textrm{diag }$\left( 1,-1,-1,-1\right) $ and we
shall always sum over repeated indices. Four-momentum operators are defined
as $p^{\mu }=i\partial ^{\mu }$, $\partial ^{\mu }=\frac{\partial }{\partial
x_{\mu }}$ where natural units have been used: $c=1$, $\hbar =1$. The
interaction is introduced via minimal coupling, $\pi ^{\mu }=p^{\mu
}-qA^{\mu }$ with a four-potential $A^{\mu }$ and a charge $q$. In the
present paper we shall consider a special classes of four-potentials obeying
the condition:%
\begin{equation}
\left[ \pi ^{0}\pm \pi ^{3},\pi ^{1}\pm i\pi ^{2}\right] =0
\label{specialA1}
\end{equation}%
where $\left[ X,Y\right] =XY-YX$ is a commutator. The condition (\ref%
{specialA1}) is fulfilled in the abelian case for%
\begin{equation}
A^{\mu }=A^{\mu }\left( x^{0},x^{3}\right) ,\ A^{i}=A^{i}\left(
x^{1},x^{2}\right) ,\ \mu =0,3,\ i=1,2.  \label{specialA2}
\end{equation}%
This is the case of longitudinal potentials, nonstandard but Lorentz
covariant, for which several exact solutions of the Dirac equation were
found \cite{Bagrov1990}.

The Dirac equation is a relativistic quantum mechanical wave equation
formulated by Paul Dirac in 1928 providing a description of elementary spin $%
\frac{1}{2}$ particles, such as electrons and quarks, consistent with both
the principles of quantum mechanics and the theory of special relativity 
\cite{Dirac1928a,Dirac1928b}. The Dirac equation is \cite%
{Bjorken1964,Berestetskii1971,Thaller1992}: 
\begin{equation}
\gamma ^{\mu }p_{\mu }\Psi =m\Psi ,  \label{Dirac1}
\end{equation}%
where $m$ is the rest mass of the elementary particle. The $\gamma $'s are $%
4\times 4$ anticommuting Dirac matrices: $\gamma ^{\mu }\gamma ^{\nu
}+\gamma ^{\nu }\gamma ^{\mu }=2g^{\mu \nu }I$ where $I$ is a unit matrix.
In the spinor representation of the Dirac matrices we have: 
\begin{equation}
\gamma ^{0}=\left( 
\begin{array}{cc}
0 & \sigma ^{0} \\ 
\sigma ^{0} & 0%
\end{array}%
\right) ,\ \gamma ^{j}=\left( 
\begin{array}{cc}
0 & -\sigma ^{j} \\ 
\sigma ^{j} & 0%
\end{array}%
\right) \ \left( j=1,2,3\right) ,  \label{gamma}
\end{equation}%
where $\sigma ^{j}$ are the Pauli matrices, $\sigma ^{0}$\ is the unit
matrix. The wave function is a bispinor, i.e. consists of $2$ two-component
spinors $\xi $, $\eta $: $\Psi =\left( \xi ,\eta \right) ^{T}$ where $^{T}$
denotes transposition of a matrix.

\section{\label{Splitting}Splitting the Dirac equation in longitudinal
external fields}

The Dirac equation (\ref{Dirac1}) can be written in spinor notation as \cite%
{Berestetskii1971}:%
\begin{equation}
\left. 
\begin{array}{r}
\pi ^{1\dot{1}}\eta _{\dot{1}}+\pi ^{1\dot{2}}\eta _{\dot{2}}=m\xi ^{1} \\ 
\pi ^{2\dot{1}}\eta _{\dot{1}}+\pi ^{2\dot{2}}\eta _{\dot{2}}=m\xi ^{2} \\ 
\pi _{1\dot{1}}\xi ^{1}+\pi _{2\dot{1}}\xi ^{2}=m\eta _{\dot{1}} \\ 
\pi _{1\dot{2}}\xi ^{1}+\pi _{2\dot{2}}\xi ^{2}=m\eta _{\dot{2}}%
\end{array}%
\right\} ,  \label{Dirac2}
\end{equation}%
where $\pi ^{A\dot{B}}$ is given by:%
\begin{equation}
\pi ^{A\dot{B}}=\left( 
\begin{array}{cc}
\pi ^{0}+\pi ^{3} & \pi ^{1}-i\pi ^{2} \\ 
\pi ^{1}+i\pi ^{2} & \pi ^{0}-\pi ^{3}%
\end{array}%
\right) ,  \label{4vector-spinor}
\end{equation}%
and $\pi _{1\dot{1}}=\pi ^{2\dot{2}}$, $\pi _{1\dot{2}}=-\pi ^{2\dot{1}}$, $%
\pi _{2\dot{1}}=-\pi ^{1\dot{2}}$, $\pi _{2\dot{2}}=\pi ^{1\dot{1}}$ (for
details of the spinor calculus reader should consult \cite%
{Berestetskii1971,MTW1973,Corson1953}). Obviously, due to relations between
components of $\pi ^{A\dot{B}}$ and $\pi _{C\dot{D}}$ the equation (\ref%
{Dirac2}) can be rewritten in terms of components of $\pi ^{A\dot{B}}$ only.
Eqn. (\ref{Dirac2}) corresponds to (\ref{Dirac1}) in the spinor
representation of $\gamma $ matrices and $\Psi =\left( \xi ^{1},\xi
^{2},\eta _{\dot{1}},\eta _{\dot{2}}\right) ^{T}$. We assume here that we
deal with four-potentials fulfilling condition (\ref{specialA1}).

In this Section we shall investigate a possibility of finding subsolutions
of the Dirac equation in longitudinal external field, analogous to
subsolutions found for the free Dirac equation in (\cite{Okninski2011}). For 
$m\neq 0$ we can define new quantities:%
\begin{eqnarray}
\pi ^{1\dot{1}}\eta _{\dot{1}} &=&m\xi _{(1)}^{1},\quad \pi ^{1\dot{2}}\eta
_{\dot{2}}=m\xi _{(2)}^{1},  \label{def1} \\
\pi ^{2\dot{1}}\eta _{\dot{1}} &=&m\xi _{(1)}^{2},\quad \pi ^{2\dot{2}}\eta
_{\dot{2}}=m\xi _{(2)}^{2},  \label{def2}
\end{eqnarray}%
where we have:%
\begin{eqnarray}
\xi _{(1)}^{1}+\xi _{(2)}^{1} &=&\xi ^{1},  \label{def3} \\
\xi _{(1)}^{2}+\xi _{(2)}^{2} &=&\xi ^{2}.  \label{def4}
\end{eqnarray}%
In spinor notation $\xi _{(1)}^{1}=\psi _{\dot{1}}^{1\dot{1}}$, $\xi
_{(2)}^{1}=\psi _{\dot{2}}^{1\dot{2}}$, $\xi _{(1)}^{2}=\psi _{\dot{1}}^{2%
\dot{1}}$, $\xi _{(2)}^{2}=\psi _{\dot{2}}^{2\dot{2}}$.

The Dirac equations (\ref{Dirac2}) can be now written with help of Eqns. (%
\ref{def3}), (\ref{def4}) as (we are now using components $\pi ^{A\dot{B}}$
throughout):%
\begin{equation}
\left. 
\begin{array}{r}
\pi ^{1\dot{1}}\eta _{\dot{1}}=m\xi _{(1)}^{1} \\ 
\pi ^{1\dot{2}}\eta _{\dot{2}}=m\xi _{(2)}^{1} \\ 
\pi ^{2\dot{1}}\eta _{\dot{1}}=m\xi _{(1)}^{2} \\ 
\pi ^{2\dot{2}}\eta _{\dot{2}}=m\xi _{(2)}^{2} \\ 
\pi ^{2\dot{2}}\left( \xi _{(1)}^{1}+\xi _{(2)}^{1}\right) -\pi ^{1\dot{2}%
}\left( \xi _{(1)}^{2}+\xi _{(2)}^{2}\right) =m\eta _{\dot{1}} \\ 
-\pi ^{2\dot{1}}\left( \xi _{(1)}^{1}+\xi _{(2)}^{1}\right) +\pi ^{1\dot{1}%
}\left( \xi _{(1)}^{2}+\xi _{(2)}^{2}\right) =m\eta _{\dot{2}}%
\end{array}%
\right\} .  \label{Dirac3}
\end{equation}%
It follows from Eqns.(\ref{def1}), (\ref{def2}) \emph{and} (\ref{specialA1})
that the following identities hold:%
\begin{eqnarray}
\pi ^{2\dot{1}}\xi _{(1)}^{1} &=&\pi ^{1\dot{1}}\xi _{(1)}^{2},  \label{id1a}
\\
\pi ^{2\dot{2}}\xi _{(2)}^{1} &=&\pi ^{1\dot{2}}\xi _{(2)}^{2}.  \label{id2a}
\end{eqnarray}%
Taking into account the identities (\ref{id1a}), (\ref{id2a}) we can
decouple Eqns. (\ref{Dirac3}) and write it as a system of the following two
equations:%
\begin{equation}
\left. 
\begin{array}{r}
\pi ^{1\dot{1}}\eta _{\dot{1}}=m\xi _{(1)}^{1} \\ 
\pi ^{2\dot{1}}\eta _{\dot{1}}=m\xi _{(1)}^{2} \\ 
\pi ^{2\dot{2}}\xi _{(1)}^{1}-\pi ^{1\dot{2}}\xi _{(1)}^{2}=m\eta _{\dot{1}}%
\end{array}%
\right\} ,  \label{constituent1}
\end{equation}%
\begin{equation}
\left. 
\begin{array}{r}
\pi ^{1\dot{2}}\eta _{\dot{2}}=m\xi _{(2)}^{1} \\ 
\pi ^{2\dot{2}}\eta _{\dot{2}}=m\xi _{(2)}^{2} \\ 
-\pi ^{2\dot{1}}\xi _{(2)}^{1}+\pi ^{1\dot{1}}\xi _{(2)}^{2}=m\eta _{\dot{2}}%
\end{array}%
\right\} .  \label{constituent2}
\end{equation}%
System of equations (\ref{constituent1}), (\ref{constituent2}), is
equivalent to the Dirac equation (\ref{Dirac2}) if the definitions (\ref%
{def3}), (\ref{def4}) are invoked.

Due to the identities (\ref{id1a}), (\ref{id2a}) equations (\ref%
{constituent1}), (\ref{constituent2}) can be cast into form: 
\begin{equation}
\left( 
\begin{array}{cccc}
0 & 0 & \pi ^{1\dot{1}} & \pi ^{1\dot{2}} \\ 
0 & 0 & \pi ^{2\dot{1}} & \pi ^{2\dot{2}} \\ 
\pi ^{2\dot{2}} & -\pi ^{1\dot{2}} & 0 & 0 \\ 
-\pi ^{2\dot{1}} & \pi ^{1\dot{1}} & 0 & 0%
\end{array}%
\right) \hspace{-0.03in}\left( 
\begin{array}{c}
\xi _{(1)}^{1} \\ 
\xi _{(1)}^{2} \\ 
\eta _{\dot{1}} \\ 
0%
\end{array}%
\right) =m\left( 
\begin{array}{c}
\xi _{(1)}^{1} \\ 
\xi _{(1)}^{2} \\ 
\eta _{\dot{1}} \\ 
0%
\end{array}%
\right) ,  \label{A-D}
\end{equation}%
\begin{equation}
\left( 
\begin{array}{cccc}
0 & 0 & \pi ^{2\dot{2}} & \pi ^{2\dot{1}} \\ 
0 & 0 & \pi ^{1\dot{2}} & \pi ^{1\dot{1}} \\ 
\pi ^{1\dot{1}} & -\pi ^{2\dot{1}} & 0 & 0 \\ 
-\pi ^{1\dot{2}} & \pi ^{2\dot{2}} & 0 & 0%
\end{array}%
\right) \hspace{-0.03in}\left( 
\begin{array}{c}
\xi _{(2)}^{2} \\ 
\xi _{(2)}^{1} \\ 
\eta _{\dot{2}} \\ 
0%
\end{array}%
\right) =m\left( 
\begin{array}{c}
\xi _{(2)}^{2} \\ 
\xi _{(2)}^{1} \\ 
\eta _{\dot{2}} \\ 
0%
\end{array}%
\right) .  \label{B-D}
\end{equation}

Let us consider Eqn. (\ref{A-D}). It can be written as:%
\begin{equation}
\gamma ^{\mu }\pi _{\mu }P_{4}\Psi _{\left( 1\right) }=mP_{4}\Psi _{\left(
1\right) },  \label{SUSY1}
\end{equation}%
where $P_{4}$ is the projection operator, $P_{4}=$ \textrm{diag }$\left(
1,1,1,0\right) $ in the spinor representation of the Dirac matrices and $%
\Psi _{\left( 1\right) }=\left( \xi _{(1)}^{1},\xi _{(1)}^{2},\eta _{\dot{1}%
},\eta _{\dot{2}}\right) ^{T}$. There are also other projection operators
which lead to analogous three component equations, $P_{1}=$ \textrm{diag }$%
\left( 0,1,1,1\right) $, $P_{2}=$ \textrm{diag }$\left( 1,0,1,1\right) $, $%
P_{3}=$ \textrm{diag }$\left( 1,1,0,1\right) $. Acting from the left on (\ref%
{SUSY1}) with $P_{4}$ and $\left( 1-P_{4}\right) $ we obtain two equations: 
\begin{eqnarray}
P_{4}\left( \gamma _{\mu }^{\mu }\pi \right) P_{4}\Psi _{\left( 1\right) }
&=&mP_{4}\Psi _{\left( 1\right) },  \label{SUSY2a} \\
\left( 1-P_{4}\right) \left( \gamma _{\mu }^{\mu }\pi \right) P_{4}\Psi
_{\left( 1\right) } &=&0.  \label{SUSY2b}
\end{eqnarray}%
In the spinor representation of $\gamma ^{\mu }$ matrices Eqn.(\ref{SUSY2a})
is equivalent to (\ref{constituent1}) while Eqn.(\ref{SUSY2b}) is equivalent
to the identity (\ref{id1a}), respectively. Now, the operator $P_{4}$ can be
written as $P_{4}=\frac{1}{4}\left( 3\mathbf{+}\gamma ^{5}-\gamma ^{0}\gamma
^{3}+i\gamma ^{1}\gamma ^{2}\right) $ where $\gamma ^{5}=i\gamma ^{0}\gamma
^{1}\gamma ^{2}\gamma ^{3}$ (similar formulae can be given for other
projection operators $P_{1},P_{2},P_{3}$, see \cite{Corson1953} where
another convention for $\gamma ^{\mu }$ matrices was however used). It thus
follows that Eqn. (\ref{SUSY1}) is given representation independent form. We
note that Eqn. (\ref{SUSY1}) is Lorentz covariant and equation (\ref{B-D})
can be obtained from (\ref{A-D})\ via charge conjugation, see Ref. \cite%
{Okninski2011} where subsolutions of form (\ref{SUSY1}) were obtained for
the free Dirac equation.

Let us note finally, that Eqn. (\ref{B-D}) can be alternatively written as%
\begin{equation}
\gamma ^{\mu }\pi _{\mu }P_{3}\Psi _{\left( 2\right) }=mP_{3}\Psi _{\left(
2\right) },  \label{SUSY2}
\end{equation}%
where $\Psi _{\left( 2\right) }=\left( \xi _{(2)}^{1},\xi _{(2)}^{2},\eta _{%
\dot{1}},\eta _{\dot{2}}\right) ^{T}$, $P_{3}=\frac{1}{4}\left( 3\mathbf{+}%
\gamma ^{5}+\gamma ^{0}\gamma ^{3}-i\gamma ^{1}\gamma ^{2}\right) $, note
that $\Psi =P_{4}\Psi _{\left( 1\right) }+P_{3}\Psi _{\left( 2\right) }$.

\section{\label{Separation}Separation of variables in subsolutions equations}

It is possible to separate variables in Eqns. (\ref{constituent1}), (\ref%
{constituent2}) exactly as described in \cite{Bagrov1990}. Substituting $\xi
_{(1)}^{1}$ and $\xi _{(1)}^{2}$ from the first two equations into the third
in Eqn. (\ref{constituent1}) we get:%
\begin{equation}
\pi ^{2\dot{2}}\pi ^{1\dot{1}}\eta _{\dot{1}}-\pi ^{1\dot{2}}\pi ^{2\dot{1}%
}\eta _{\dot{1}}=m^{2}\eta _{\dot{1}}.  \label{eta1}
\end{equation}%
Taking into account definition of $\pi ^{A\dot{B}}$ and property (\ref%
{specialA2}) we obtain:%
\begin{equation}
\left( \pi _{\mu }\pi ^{\mu }+iqE\left( x^{0},x^{3}\right) +qH\left(
x^{1},x^{2}\right) \right) \eta _{\dot{1}}=m^{2}\eta _{\dot{1}}  \label{eta2}
\end{equation}%
where $E=\partial _{0}A_{3}-\partial _{3}A_{0}$, $H=\partial
_{2}A_{1}-\partial _{1}A_{2}$.

To achieve separation of variables we put:

\begin{eqnarray}
\eta _{\dot{1}}\left( x\right) &=&\varphi _{\dot{1}}\left(
x^{0},x^{3}\right) \psi _{\dot{1}}\left( x^{1},x^{2}\right) ,  \label{V1} \\
\xi _{(1)}^{1} &=&\alpha _{\dot{1}}\left( x^{0},x^{3}\right) \psi _{\dot{1}%
}\left( x^{1},x^{2}\right) ,  \label{V2} \\
\xi _{(1)}^{2} &=&\varphi _{\dot{1}}\left( x^{0},x^{3}\right) \beta _{\dot{1}%
}\left( x^{1},x^{2}\right) .  \label{V3}
\end{eqnarray}

We now substitute Eqn. (\ref{V1}) into (\ref{eta2}) to get: 
\begin{subequations}
\label{S12}
\begin{eqnarray}
\left( \left( \pi ^{0}\right) ^{2}-\left( \pi ^{3}\right) ^{2}+iqE\left(
x^{0},x^{3}\right) \right) \varphi _{\dot{1}}\left( x^{0},x^{3}\right) \!
&=&\!\left( m^{2}+\lambda _{\dot{1}}^{2}\right) \varphi _{\dot{1}}\left(
x^{0},x^{3}\right)   \label{S1} \\
\left( \left( \pi ^{1}\right) ^{2}+\left( \pi ^{2}\right) ^{2}-qH\left(
x^{1},x^{2}\right) \right) \psi _{\dot{1}}\left( x^{1},x^{2}\right) \!
&=&\!\lambda _{\dot{1}}^{2}\psi _{\dot{1}}\left( x^{1},x^{2}\right) 
\label{S2}
\end{eqnarray}%
where $\lambda _{\dot{1}}^{2}$\ is the separation constant and we note that
equations. (\ref{S1}), (\ref{S2}) are analogous to Eqns. (12.15), (12.19) in 
\cite{Bagrov1990}.

Combining now Eqn. (\ref{S1}) with the first of Eqns. (\ref{constituent1})
we obtain: 
\end{subequations}
\begin{subequations}
\label{F12a}
\begin{eqnarray}
\left( \pi ^{0}+\pi ^{3}\right) \varphi _{\dot{1}}\left( x^{0},x^{3}\right) 
&=&m\alpha _{\dot{1}}\left( x^{0},x^{3}\right) ,  \label{F1a} \\
\left( \pi ^{0}-\pi ^{3}\right) \alpha _{\dot{1}}\left( x^{0},x^{3}\right) 
&=&\frac{m^{2}+\lambda _{\dot{1}}^{2}}{m}\varphi _{\dot{1}}\left(
x^{0},x^{3}\right) ,  \label{F2a}
\end{eqnarray}%
while combining Eqn. (\ref{S2}) with the second of Eqns. (\ref{constituent1}
) we get equations: 
\end{subequations}
\begin{subequations}
\label{B12}
\begin{eqnarray}
\left( \pi ^{1}-i\pi ^{2}\right) \psi _{\dot{1}}\left( x^{1},x^{2}\right) 
&=&m\beta _{\dot{1}}\left( x^{1},x^{2}\right) ,  \label{B1} \\
\left( \pi ^{1}+i\pi ^{2}\right) \beta _{\dot{1}}\left( x^{1},x^{2}\right) 
&=&\frac{\lambda _{\dot{1}}^{2}}{m}\psi _{\dot{1}}\left( x^{1},x^{2}\right) ,
\label{B2}
\end{eqnarray}%
which can be also written as the Pauli equation: 
\end{subequations}
\begin{equation}
\left[ \left( \left( \pi ^{1}\right) ^{2}+\left( \pi ^{2}\right) ^{2}\right)
\sigma ^{0}-qH\left( x^{1},x^{2}\right) \sigma ^{3}\right] \left( 
\begin{array}{c}
\psi _{\dot{1}} \\ 
\beta _{\dot{1}}%
\end{array}%
\right) =\lambda _{\dot{1}}^{2}\left( 
\begin{array}{c}
\psi _{\dot{1}} \\ 
\beta _{\dot{1}}%
\end{array}%
\right) .  \label{Pauli1}
\end{equation}

Finally, after the rescaling $\alpha _{\dot{1}}\left( x^{0},x^{3}\right) =%
\sqrt{1+\frac{\lambda _{\dot{1}}^{2}}{m^{2}}}\tilde{\alpha}_{\dot{1}}\left(
x^{0},x^{3}\right) $, equations (\ref{F1a}), (\ref{F2a}) are converted to $%
2D $ Dirac equation: 
\begin{subequations}
\label{F12b}
\begin{eqnarray}
\left( \pi ^{0}+\pi ^{3}\right) \varphi _{\dot{1}}\left( x^{0},x^{3}\right)
&=&\tilde{m}\tilde{\alpha}_{\dot{1}}\left( x^{0},x^{3}\right) ,  \label{F1b}
\\
\left( \pi ^{0}-\pi ^{3}\right) \tilde{\alpha}_{\dot{1}}\left(
x^{0},x^{3}\right) &=&\tilde{m}\varphi _{\dot{1}}\left( x^{0},x^{3}\right) ,
\label{F2b}
\end{eqnarray}%
with effective mass $\tilde{m}=\sqrt{m^{2}+\lambda _{\dot{1}}^{2}}$.

The same procedure applied to Eqn. (\ref{constituent2}) yields the equation
for $\eta _{\dot{2}}$: 
\end{subequations}
\begin{equation}
\left( \pi _{\mu }\pi ^{\mu }-iqE\left( x^{0},x^{3}\right) -qH\left(
x^{1},x^{2}\right) \right) \eta _{\dot{2}}=m^{2}\eta _{\dot{2}}.
\label{eta3}
\end{equation}%
Carrying out separation of variables we get:

\begin{subequations}
\label{S34}
\begin{eqnarray}
\left( \left( \pi ^{0}\right) ^{2}-\left( \pi ^{3}\right) ^{2}-iqE\left(
x^{0},x^{3}\right) \right) \!\varphi _{\dot{2}}\left( x^{0},x^{3}\right) 
&=&\!\left( m^{2}+\lambda _{\dot{2}}^{2}\right) \varphi _{\dot{2}}\left(
x^{0},x^{3}\right)   \label{S3} \\
\left( \left( \pi ^{1}\right) ^{2}+\left( \pi ^{2}\right) ^{2}+qH\left(
x^{1},x^{2}\right) \right) \!\psi _{\dot{2}}\left( x^{1},x^{2}\right) 
&=&\!\lambda _{\dot{2}}^{2}\psi _{\dot{2}}\left( x^{1},x^{2}\right) 
\label{S4}
\end{eqnarray}%
and 
\end{subequations}
\begin{subequations}
\label{B34}
\begin{eqnarray}
\left( \pi ^{1}+i\pi ^{2}\right) \psi _{\dot{2}}\left( x^{1},x^{2}\right) 
&=&m\beta _{\dot{2}}\left( x^{1},x^{2}\right) ,  \label{B3} \\
\left( \pi ^{1}-i\pi ^{2}\right) \beta _{\dot{2}}\left( x^{1},x^{2}\right) 
&=&\frac{\lambda _{\dot{2}}^{2}}{m}\psi _{\dot{2}}\left( x^{1},x^{2}\right) ,
\label{B4}
\end{eqnarray}%
which is written as the Pauli equation 
\end{subequations}
\begin{equation}
\left[ \left( \left( \pi ^{1}\right) ^{2}+\left( \pi ^{2}\right) ^{2}\right)
\sigma ^{0}+qH\left( x^{1},x^{2}\right) \sigma ^{3}\right] \left( 
\begin{array}{c}
\psi _{\dot{2}} \\ 
\beta _{\dot{2}}%
\end{array}%
\right) =\lambda _{\dot{2}}^{2}\left( 
\begin{array}{c}
\psi _{\dot{2}} \\ 
\beta _{\dot{2}}%
\end{array}%
\right) .  \label{Pauli2}
\end{equation}%
and the Dirac equation:

\begin{subequations}
\label{F34b}
\begin{eqnarray}
\left( \pi ^{0}-\pi ^{3}\right) \varphi _{\dot{2}}\left( x^{0},x^{3}\right)
&=&\hat{m}\hat{\alpha}_{\dot{2}}\left( x^{0},x^{3}\right) ,  \label{F3b} \\
\left( \pi ^{0}+\pi ^{3}\right) \hat{\alpha}_{\dot{2}}\left(
x^{0},x^{3}\right) &=&\hat{m}\varphi _{\dot{2}}\left( x^{0},x^{3}\right) ,
\label{F4b}
\end{eqnarray}%
with effective mass $\hat{m}=\sqrt{m^{2}+\lambda _{\dot{2}}^{2}}$ and $%
\alpha _{\dot{2}}\left( x^{0},x^{3}\right) =\sqrt{1+\frac{\lambda _{\dot{2}%
}^{2}}{m^{2}}}\hat{\alpha}_{\dot{2}}\left( x^{0},x^{3}\right) $ where the
following definitions were used: 
\end{subequations}
\begin{eqnarray}
\eta _{\dot{2}}\left( x\right) &=&\varphi _{\dot{2}}\left(
x^{0},x^{3}\right) \psi _{\dot{2}}\left( x^{1},x^{2}\right) ,  \label{V4} \\
\xi _{(2)}^{1} &=&\alpha _{\dot{2}}\left( x^{0},x^{3}\right) \psi _{\dot{2}%
}\left( x^{1},x^{2}\right) ,  \label{V5} \\
\xi _{(2)}^{2} &=&\varphi _{\dot{2}}\left( x^{0},x^{3}\right) \beta _{\dot{2}%
}\left( x^{1},x^{2}\right) .  \label{V6}
\end{eqnarray}

\section{\label{Discussion}Discussion}

We have shown that the Dirac equation in longitudinal external fields can be
splitted into two subsolutions equations, (\ref{SUSY1}), (\ref{SUSY2}),
which are Dirac equations with built-in projection operators. These
covariant equations describe a particle and an antiparticle, respectively,
as follows from Eqns. (\ref{eta2}), (\ref{eta3}), see also \cite%
{Okninski2011} and references therein. Next we have separated variables in
subsolutions equations obtaining Dirac equations in $1+1$ dimensions (\ref%
{F12b}), (\ref{F34b}) and $2D$ Pauli equations (\ref{Pauli1}), (\ref{Pauli2}%
). These equations lead to the picture of a particle\thinspace /\thinspace
antiparticle moving with effective mass along the $x^{3}$ axis in external
electric field, transversal degrees of freedom interacting with external
magnetic field.


\begin{thebibliography}{99}
\bibitem{Jackiw1991} R.~Jackiw, V.P. Nair, Phys. \ Rev. \textbf{D43}, 1933
(1991)

\bibitem{Plyushchay1991} M.S. Plyushchay, Phys. Lett. \textbf{B273}, 250
(1991)

\bibitem{Horvathy2010a} P.~Horv\'{a}thy, M.~Plyushchay, M.~Valenzuela,
Annals of Physics \textbf{325}, 1931 (2010)

\bibitem{Horvathy2010b} P.~Horv\'{a}thy, M.~Plyushchay, M.~Valenzuela, Phys.
\ Rev. \textbf{D81}, 127701 (2010)

\bibitem{Horvathy2010c} P.~Horv\'{a}thy, M.~Plyushchay, M.~Valenzuela, J.
Math. Phys. \textbf{51}, 092108 (2010)

\bibitem{Horvathy2008} P.~Horv\'{a}thy, M.~Plyushchay, M.~Valenzuela, Phys.
Rev. \textbf{D77}, 025017 (2008)

\bibitem{Simulik1998} V.M Simulik, I.Yu. Krivsky, Adv. Appl. Clifford Algebras 
\textbf{8}, 69 (1998)

\bibitem{Simulik2011} V.M Simulik, I.Yu. Krivsky, Phys. Lett. \textbf{A 375}, 2479 (2011)

\bibitem{Okninski2011} A.~Okninski, Int. J. Theor. Phys \textbf{50}, 729
(2011)

\bibitem{Bagrov1990} V. G. Bagrov, D. M. Gitman, \emph{Exact Solutions of
Relativistic Wave Equations,} (Kluver Academic Publishers, 1990)

\bibitem{Dirac1928a} P.~Dirac, Proc. Roy. Soc. London \textbf{A117}, 610
(1928)

\bibitem{Dirac1928b} P.~Dirac, Proc. Roy. Soc. London \textbf{A118}, 351
(1928)

\bibitem{Bjorken1964} J.~Bjorken, S.~Drell, \emph{Relativistic Quantum
Mechanics} (McGraw-Hill, Inc., 1964)

\bibitem{Berestetskii1971} V.~Berestetskii, E.~Lifshits, L.~Pitaevskii, 
\emph{Relativistic Quantum Theory} (Pergamon, 1974)

\bibitem{Thaller1992} B.~Thaller, \emph{The Dirac Equation}
(Springer-Verlag, 1992)

\bibitem{Duffin1938} R.~Duffin, Phys.\ Rev. \textbf{54}, 1114 (1938)

\bibitem{Kemmer1939} N.~Kemmer, Proc. Roy. Soc. (London) \textbf{A173}, 91
(1939)

\bibitem{Kemmer1943} N.~Kemmer, Proc. Camb. Phil. Soc. \textbf{39}, 189
(1943)

\bibitem{Petiau1936} G.~Petiau, Mem. Acad. Sci. Roy. Belgique \textbf{16},
No. 2, 1 (1936)

\bibitem{Proca1936a} A.~Proca, J. Phys. Radium \textbf{7}, 347 (1936)

\bibitem{Proca1936b} A.~Proca, C.R. Acad. Sci. Paris \textbf{1366}, 347
(1936)

\bibitem{Lanczos1929} C.~Lanczos, Z. Phys. \textbf{1366}, 447, 474, 484
(1929)

\bibitem{Bogush2007} A.~Bogush, V.~Kisel, N.~Tokarevskaya, V.~Red'kov, Ann.
Fond. Louis de Broglie \textbf{32}, 355 (2007)

\bibitem{MTW1973} C.~Misner, K.~Thorne, J.~Wheeler, \emph{Gravitation} (W.H.
Freeman and Company, 1973)

\bibitem{Corson1953} E.~Corson, \emph{Introduction to Tensors, Spinors, and
Relativistic Wave-Equations (Relation Structure)} (Blackie and Son Limited,
1953)
\end{thebibliography}
\end{document}